\documentclass{article}
\usepackage{graphicx}
\usepackage{newcent} 
\usepackage{mathptmx}      
\usepackage{fancyhdr}
\usepackage{hyperref}
\usepackage{cite}
\usepackage[a4paper,top=3cm,bottom=3.3cm,left=3cm,right=3cm,footskip=1.5cm,truedimen,headheight=17pt]{geometry}
\hypersetup{colorlinks=true,citecolor=red,urlcolor=blue}
\begin{document}
\title{Slightly imbalanced system of a few attractive fermions \\ in a one-dimensional harmonic trap}%
\author{Tomasz Sowi\'nski \\
           \small Institute of Physics of the Polish Academy of Sciences \\
           \small Al. Lotnik\'ow 32/46, 02-668 Warsaw, Poland\\
           \small \texttt{Tomasz.Sowinski@ifpan.edu.pl}    
              }
\date{\it In Honor of Professor Maciej Lewenstein on his 60th birthday}

\maketitle
\thispagestyle{fancy}
\begin{abstract}
The ground-state properties of the two-flavored mixture of a few attractive fermions confined in a one-dimensional harmonic trap is studied. It is shown that for slightly imbalanced system the pairing between fermions of opposite spins has completely different  features that in the balanced case. The fraction of correlated pairs is suppressed by the presence of additional particle and another uncorrelated two-body orbital dominates in the ground-state of the system. \\[10pt]
This is the author's version accepted for publication. \\[-2pt] 
The final publication is available at Springer via DOI: \href{http://www.dx.doi.org/10.1007/s00601-015-1017-5}{10.1007/s00601-015-1017-5}
\end{abstract}  

\section{Introduction}
Quantum few-body systems are believed to form some kind of bridge between one- and two-body physics, well described with standard methods of quantum mechanics, and many-body world where collective properties of quantum particles have to be taken into account in the framework of statistical mechanics \cite{Blume,ZinnerRev}. In nowadays experiments on ultra-cold atoms it is possible to engineer, to control, and to perform measurements on few-body systems in many different schemes \cite{Serwane2011,Wenz2013few,Haller2009,Zurn}. This amazing experimental progress has inspired theoreticians to reformulate old and to formulate new questions on the properties of a mesoscopic number of quantum particles \cite{SowinskiGrass,Blume2013,ZinnerSep,ZinnerMulti,Cui,Blume2014,Harshman,ZinnerEPJ,Sowinski2015,DAmico2015,Pecak}. One of them is the question on existence of pairing in attractive systems of a few, not-uniformly confined fermions of two flavors \cite{Zurn}. Partial answer for a one-dimensional system was given recently \cite{Sowinski2015,DAmico2015}.  It was shown that for balanced system, i.e. when numbers of fermions in both flavors are equal, in the presence of strong attraction between fermions, the fraction of the correlated pairs with opposite spins starts to dominate in the system. The effective two-particle orbital of paired fermions, in contrast to other orbitals of the two-particle reduced density matrix, has many properties of the Cooper orbital known from the theory of superconductivity \cite{Cooper,BCSTheory}. In this way a theoretical link to the system in the thermodynamic limit was established \cite{Sowinski2015}. However, the analysis performed in \cite{Sowinski2015} was not extended to imbalanced systems, i.e. cases when numbers of fermions in both flavors are different. Since properties of a one-dimensional system can be completely different than its properties in higher dimensions, therefore the question on pairing in imbalanced system become even more important. To fill this gap, in this article I report first, surprising results on the properties of the two-flavored mixture of fermions (confined in a one-dimensional trap) when one additional fermion of chosen flavor is present in the system. 

\section{The Model}
I consider an isolated system of a few ultra-cold fermions of mass $m$ belonging to two, fundamentally  distinguishable, flavors confined in a one-dimensional harmonic trap of frequency $\Omega$ and interacting via short-range $\delta$-like potential. In the second quantization formalism the Hamiltonian has a form \cite{SowinskiGrass,Sowinski2015}
\begin{equation} \label{Ham1}
\widehat{\cal H} = \sum_\sigma\int\!\mathrm{d}x\,\,\widehat\Psi_\sigma^\dagger(x)\left[-\frac{\hbar^2}{2m}\frac{\mathrm{d}^2}{\mathrm{d}x^2}+\frac{m\Omega^2}{2}x^2\right] \widehat\Psi_\sigma(x) + g\int\!\mathrm{d}x\,\, 
\widehat\Psi_\uparrow^\dagger(x)\widehat\Psi_\downarrow^\dagger(x)\widehat\Psi_\downarrow(x)\widehat\Psi_\uparrow(x).
\end{equation}
The field operator $\widehat\Psi_\sigma(x)$ annihilates a fermion with spin $\sigma$ at a position $x$. Parameter $g$ denotes the effective interaction constant between fermions of opposite spins which can be controlled experimentally.  Since we assume that fermions of opposite spins are fully distinguishable the corresponding field operators do commute, i.e. $\left[\widehat\Psi_\uparrow(x),\widehat\Psi_\downarrow(x')\right]=\left[\widehat\Psi_\uparrow(x),\widehat\Psi^\dagger_\downarrow(x')\right]=0$. Of course, the field operators of the same spins fulfill standard anti-commutation relations $\left\{\widehat\Psi_\sigma(x),\widehat\Psi_\sigma(x')\right\}=0$ and $\left\{\widehat\Psi_\sigma(x),\widehat\Psi^\dagger_\sigma(x')\right\}=\delta(x-x')$. It is quite obvious that the Hamiltonian (\ref{Ham1}) commutes with the operator of the total number of particles of given spin $\widehat{N}_\sigma = \int\!\mathrm{d}x\,\,\widehat\Psi_\sigma^\dagger(x)\widehat\Psi_\sigma(x)$. Since the Hamiltonian does not couple states with different number of fermions of given flavor, therefore one can study its properties independently in subspaces of fixed $N_\uparrow$ and $N_\downarrow$. It is worth noticing that this theoretical observation is also relevant from the experimental point of view. Indeed, in experiments with ultra-cold mixtures of fermions, two flavors are fundamentally distinguishable and superpositions of quantum states with different number of fermions are strictly forbidden. This fact can be viewed as an additional {\em superselection} principle in the model. In the following I shall assume that the numbers $N_\uparrow$ and $N_\downarrow$ are precisely known and differ by 1. Without loosing generality, I fix the relation $N_\uparrow=N_\downarrow+1$. 

\section{Exact diagonalization approach}
It was shown in \cite{Sowinski2015} that in the system described by the Hamiltonian (\ref{Ham1}) the correlated pairs appear in the system for relatively strong attraction. It means that any numerical method based on perturbative arguments fails to describe this phenomenon correctly. Therefore, the best way to study properties of the system is brut-force method originating in the direct, numerically exact diagonalization of the Hamiltonian. To perform the diagonalization I decompose the field operators $\widehat{\Psi}_\sigma(x)$ in the set of single-particle eigenstates of the harmonic oscillator $\left\{\varphi_i(x)\right\}$ cut-off on sufficiently large state $n_{max}$:
\begin{equation}
\widehat{\Psi}_\sigma(x) = \sum_{i=0}^{n_{max}} \widehat{b}_{i\sigma}\varphi_i(x).
\end{equation}
The operator $\widehat{b}_{i\sigma}$ annihilates a fermion with spin $\sigma$ in a spatial state described by the wave function $\varphi_i(x)$. With this expansion the Hamiltonian (\ref{Ham1}) is transformed to the following form:
\begin{equation} \label{Ham2}
\widehat{\cal H} = \sum_\sigma\sum_i E_i\,\widehat{b}^\dagger_{i\sigma}\widehat{b}_{i\sigma} + \sum_{ijkl} U_{ijkl} \widehat{b}^\dagger_{i\uparrow}\widehat{b}^\dagger_{j\downarrow}\widehat{b}_{k\downarrow}\widehat{b}_{l\uparrow},
\end{equation}
where $E_i=\hbar\Omega(i+\frac{1}{2})$ is the single-particle energy of $i$-th eigenstate of the harmonic confinement. Interstate interaction terms have the form:
\begin{equation}
U_{ijkl} = g \int \!\mathrm{d}x\,\, \varphi^*_i(x)\varphi^*_j(x)\varphi_k(x)\varphi_l(x).
\end{equation}
Assuming that the numbers of particles in both flavors $N_\uparrow$ and $N_\downarrow$ are fixed one can construct the Fock space of all possible distributions of particles in assumed single-particle states. Then, all matrix elements of the Hamiltonian (\ref{Ham2}) can be calculated and the resulting matrix can be diagonalized. Since the size of the matrix grows exponentially with the size of the system the method is limited to relatively small number of particles. The diagonalization is performed via the Implicitly Restarted Arnoldi Method \cite{Arnoldi}. As the result one obtains the energy of the ground-state and the decomposition of the ground-state of the system $|\mathtt{G}\rangle$ in the Fock basis defined above. 
 
At this point it is worth noticing that attractive interactions considered here are strong enough to assure that the system is far from the perturbative regime of almost non-interacting particles. However at the same time, it is still far from the region where bound molecules are formed in the system. For balanced system it is known, that in this range of interactions correlated pairs of opposite-spin fermions emerge in the system. However, the pairs cannot be viewed as bounded molecules since their size, when compared to the size of the whole many-body system, {\em increases} with interactions. In contrast, as explained in \cite{Sowinski2015}, in this range of interactions the pairs have many properties of Cooper pairs known from BCS theory \cite{Cooper,BCSTheory}. 

\section{The results}
\begin{figure}[!ht]
\centering
 \includegraphics{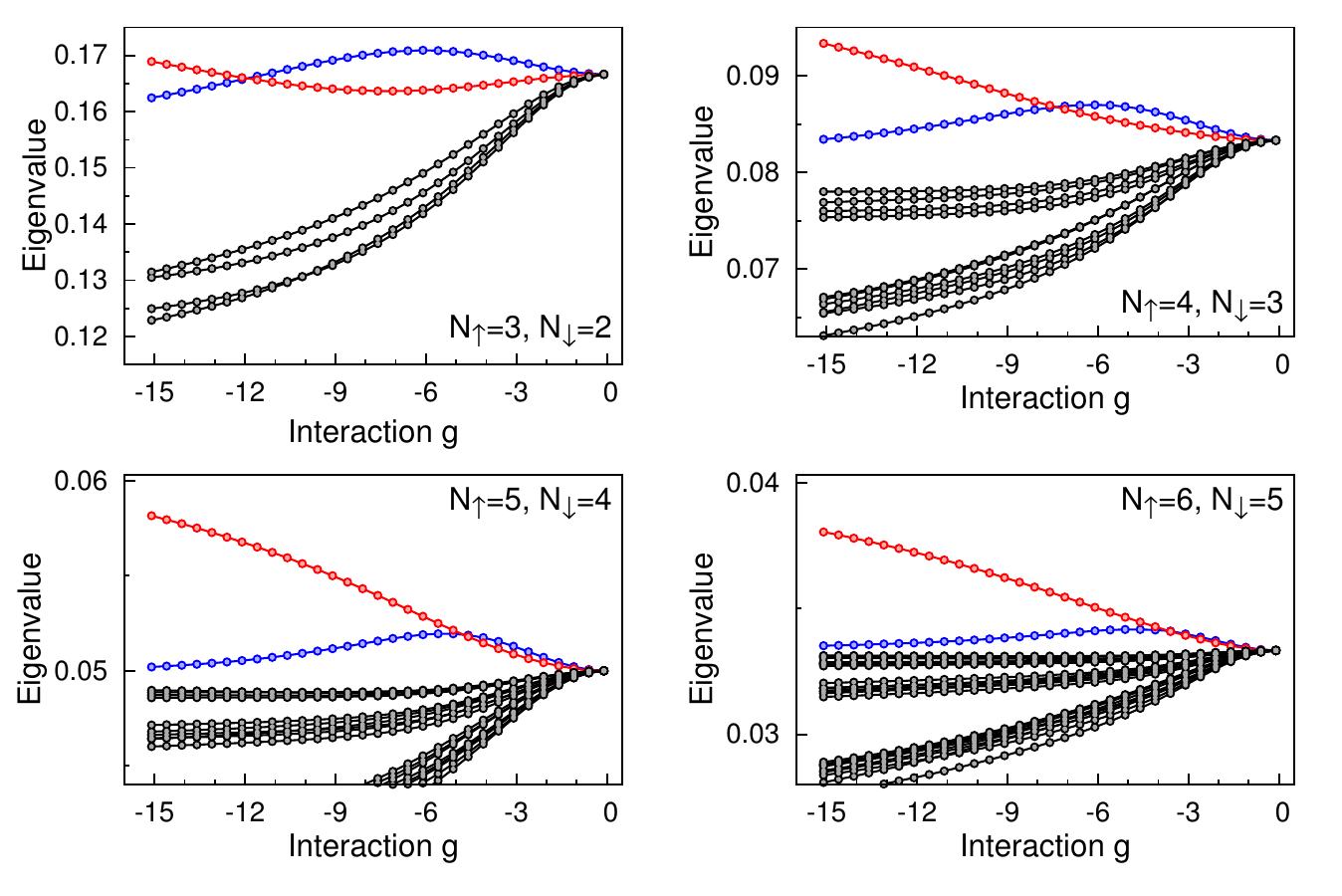}
\caption{
(color on-line) Eigenvalues of the reduced two-particle density matrix $\rho^{(2)}$ as functions of the interaction strength $g$. For noninteracting case all non-vanishing  eigenvalues are equal to $(N_\uparrow N_\downarrow)^{-1}$. For non-zero but small interaction two different orbitals start to dominate in the system. The one with the largest occupation (blue curve) manifests correlations similar to the Cooper-like pairing known from the theory of superconductivity \cite{BCSTheory}. However, for strong interactions the second orbital (free of any pairing correlations, red curve) dominates in the system solely. Note that the value of critical interaction $g_c$ for which both orbitals are equally occupied depends on the number of particles in the system.
}
\label{Fig1}      
\end{figure}

To find any tracks of possible pairing in the system, I concentrate on two-body correlations between fermions of opposite spins. Therefore, the reduced two-particle density matrix of the following form
\begin{equation}
\rho^{(2)}(x_2,x_1;x_1',x_2') = \frac{1}{N_\uparrow N_\downarrow}\langle\mathtt{G}|\widehat\Psi_\uparrow^\dagger(x_2)\widehat\Psi_\downarrow^\dagger(x_1)\widehat\Psi_\downarrow(x_1')\widehat\Psi_\uparrow(x_2')|\mathtt{G}\rangle
\end{equation}
is calculated. Then it is spectrally decomposed to its natural orbitals:
\begin{equation}
\rho^{(2)}(x_2,x_1;x_1',x_2') = \sum_n \lambda_n \Phi^*_n(x_1,x_2)\Phi_n(x_1',x_2').
\end{equation}
Technically, the decomposition is done by a diagonalization of $\rho^{(2)}$ in the harmonic oscillator representation. From the physical point of view, the eigenvalues $\{\lambda_n\}$ are understood as occupations of corresponding orbitals and they express probabilities that two opposite-spin fermions can be found in a given two-particle orbital. Whenever one of the orbitals starts to dominate in the system, the condensation of opposite-spin pair in the selected orbital occurs. In Fig. \ref{Fig1} eigenfunctions of $\rho^{(2)}$ as functions of the coupling constant $g$, for different number of fermions, are presented. As suspected for ideal gas of noninteracting fermions ($g=0$) all non-zero eigenvalues are equal to $(N_\uparrow N_\downarrow)^{-1}$. They correspond to the combinatoric number of all possible pairings of opposite spins.  The situation is different when attractive interaction is present. For small attractions, two orbitals start to dominate in the system (red and blue curve in Fig. \ref{Fig1}). However, for strong enough interaction the second orbital (red curve) is enhanced an finally it dominates in the system solely. As it is seen in Fig. 
\ref{Fig1}, the critical interaction $g_c$ at which both dominant orbitals have the same occupation strongly depends on the total number of particles in the system and it rapidly decreases for larger systems. It is worth noticing that this behavior was not present for balanced system where only one dominant orbital was present for all interactions \cite{Sowinski2015}. This observation suggests that properties of the ground-state for imbalanced system are completely different than in the case of $N_\uparrow=N_\downarrow$. 

Direct inspection on the structure of both dominant orbitals shows that they have completely different properties. It can be shown by calculating different correlation functions in these two-body wave functions. The easiest way to do this, is first to decompose orbitals to their natural representation in harmonic oscillator basis:
\begin{equation}
|\!|\Phi_n\rangle\!\rangle = \sum_{ij} \alpha_n^{ij}\,\, \widehat{b}^\dagger_{i\uparrow}\widehat{b}^\dagger_{j\downarrow}|\mathtt{vac}\rangle.
\end{equation}
The simplest correlations  which distinguish dominant orbitals expressed in this representation are:
\begin{equation}
\widehat{\cal O}_1 = \sum_{ij} \widehat{b}^\dagger_{i\uparrow}\widehat{b}^\dagger_{i\downarrow}\widehat{b}_{j\downarrow}\widehat{b}_{j\uparrow},  \qquad \widehat{\cal O}_2 = \sum_{ijk} \widehat{b}^\dagger_{i\uparrow}\widehat{b}^\dagger_{j\downarrow}\widehat{b}_{k\downarrow}\widehat{b}_{k\uparrow}.
\end{equation}
The operator $\widehat{\cal O}_1$ measures ''the mobility'' of the correlated fermionic pair. It moves correlated pairs between different single-particle levels preserving their correlations. It can be viewed as a counterpart of hopping correlation considered in lattice models \cite{Pietraszewicz}. In contrast, the expectation value of the operator $\widehat{\cal O}_2$ can be interpreted as the order parameter of the superfluid fraction, i.e. it is a counterpart of the annihilation operator of the correlated pair $\langle \Psi_\downarrow(x)\Psi_\uparrow(x)\rangle$ redefined for models with conserved number of particles (it annihilates the pair in a given state and creates uncorrelated fermions in arbitrary states). It is worth noticing that in the case of balanced system both correlations were significant only in the dominant orbital. It was interpreted as a direct manifestation of Cooper-like pairing in the system \cite{Sowinski2015,Cooper}. Similarly, for the system studied, correlations $\widehat{O}_1$ and $\widehat{O}_2$ appear also in single two-fermion orbital. However, in this case they are non-zero in the {\sl second} dominant orbital -- the orbital which dominates only in the perturbative region of small interactions (blue curve). It means that for strong interactions pairing induced by attractions is essentially suppressed and in practice cannot be observed. This result shade some fresh light to the problem of pairing induced by interactions in one-dimensional few-body systems. It suggests that, whenever system is slightly imbalanced, the Cooper mechanism is present only in the perturbative regime of interactions. Far from the single-particle picture, properties of the system are determined by processes which have no single- or two-particle description. 
 
\section{Conclusions}
I have shown that properties of two-flavored mixture of a few attractive fermions in a one-dimensional trap may crucially depend on the difference in the number of particles. Particularly, for almost balanced system, i.e. when the number of particles differ only by 1, Cooper-like pairing explored recently for balanced systems, is strongly suppressed and probably can not be observed experimentally. For small attractive forces, i.e. in the case when interactions can be treated perturbatively, the two-fermion orbital of correlated pairs dominates in the system. At the same time another two-fermion orbital, which does not manifest any pairing correlations, is occupied with almost the same probability. For strong repulsions the uncorrelated orbital dominates in the system. The result suggests that for larger difference in the number of particles between flavors a physical picture of two-particle correlations will be more complicated and it still needs theoretical exploration. 

\section{acknowledgements}
This work was supported by the (Polish) Ministry of Sciences and Higher Education, Iuventus Plus 2015-2017 Grant No. 0440/IP3/2015/73.

\end{document}